\documentstyle[epsf,12pt]{article}

\def\ave#1{\left\langle #1\right\rangle}

\begin{document}

\begin{flushright}
\today\\
\end{flushright}
\bigskip
\begin{center}
\Large
{\bf Berry-Robnik level statistics in a smooth billiard system}\\
\vspace{0.4in}
\large
Toma\v z Prosen\footnote{e-mail prosen@fiz.uni-lj.si}\\
\normalsize

\vspace{0.3in}
Department of Physics, Faculty of Mathematics and Physics,\\
University of Ljubljana, Jadranska 19, SLO-1111 Ljubljana, Slovenia
\end{center}
\vspace{0.4in}
\normalsize
{\bf Abstract.}
Berry-Robnik level spacing distribution is demonstrated clearly
in a generic quantized plane billiard for the first time.
However, this ultimate semi-classical distribution is found to be
valid only for extremely small semi-classical parameter (effective
Planck's constant) where the assumption of statistical
independence of regular and irregular levels is achieved.
For sufficiently larger semiclassical parameter
we find (fractional power-law) level repulsion with phenomenological
Brody distribution providing an adequate global fit.
\\\\
\noindent Submitted to {\bf Journal of Physics A: Math. Gen.}
\normalsize
\vspace{0.1in}
\newpage

\noindent
Energy level statistics of mixed quantum systems whose classical
dynamics is partly regular and partly chaotic have been 
intensively studied over the past decade (see \cite{PR94b} and
references therein), and this subject is still much less theoretically 
understood than the level statistics of the two extreme cases, namely  
completely chaotic (hyperbolic) systems \cite{Chaotic,Chaotic2}, 
and integrable systems \cite{Integrable}. 
However, it is believed that mixed systems, for
example hydrogen atom in strong magnetic field \cite{HRW}, 
are generic in nature, at least among dynamical systems with
few degrees of freedom.
Although Berry and Robnik have developed a semiclassical theory 
of level spacing statistics for mixed systems back in 1984 \cite{BR84}, 
there has been a lot of confusion in the literature advocating various
phenomenological models due to incompatibility of experimental or numerical 
data with the Berry-Robnik (BR) statistics (see a recent comment \cite{PR97}).
BR distribution is built on a simple and clean assumption of a {\em 
statistically independent} superposition of partial subspectra consisting 
of {\em regular} or {\em chaotic} levels 
(following an old Percivals' idea \cite{P73} of classifying the quantum 
eigenstates of mixed systems as {\em regular} or {\em chaotic}). 
The sequence of regular levels, associated to 
eigenstates whose phase space distribution functions (e.g. Wigner or Husimi) 
localize on regions of regular motion, is assumed to have Poissonian 
statistics, whereas the sequences of chaotic levels, associated with 
eigenstates whose phase space distribution functions extend over chaotic 
components of classical phase space, are assumed to have GOE (or GUE if 
antiunitary symmetry is absent) statistics of ensembles of 
Gaussian random matrices. Further, it is crucial to note that the {\em gap 
distribution} $E(S)$, the probability
that unfolded energy interval of length $S$ contains no levels,
factorizes upon independent superposition of level sequences, so
the 2-component BR distribution for a system with a single classically
chaotic component of relative measure $\rho_2$ and regular components
of complementary measure $\rho_1 = 1 - \rho_2$ reads
\begin{equation}
E_{\rm BR}(S) = E_{\rm Poisson}(\rho_1 S) E_{\rm GOE}(\rho_2 S).
\label{eq:Epr}
\end{equation}
Note that $E_{\rm Poisson}(S)=\exp(-S)$ while for
$E_{\rm GOE}(S)$ no closed-form expression exists (for the exact 
infinitely-dimensional GOE), and we have to rely on
various expansions (we recommend Pad\' e approximation published in
\cite{H91}). The more common nearest neighbour level spacing 
distribution $P(S)$ is directly related to the gap distribution, 
simply as $P_{\rm BR}(S) = d^2 E_{\rm BR}(S)/dS^2$.
\\\\
However, for the validity of semiclassical BR formula, two
conditions have to be satisfied.
(i) The regular and irregular levels should not be correlated, 
i.e. the corresponding (Wigner or Husimi) phase-space distributions 
should not overlap. This is true if the quantum resolution scale in
phase space, $\hbar^{1/2}$ (where $\hbar$ is the effective Planck's constant), 
is small enough to resolve the essential features of the structure of 
classical phase space:
$\hbar^{1/2} < $ (sizes of the main regular islands, widths of
chaotic strips penetrating through regular islands, etc).
(ii) The quantum relaxation time, i.e. the Heisenberg (break) time
$t_{\rm break} = 2\pi\hbar/\Delta E$ (where $\Delta E$ is the mean level 
spacing) should be larger than the classical ergodic time $t_{\rm erg}$ on 
the chaotic component, $ t_{\rm break} > t_{\rm erg}$.
When this is not true, one expects dynamical localization of eigenstates inside
the chaotic component \cite{PR94b,P96,BCL,CP}.

Note that the BR statistics are incompatible with level repulsion, namely
$P_{BR}(0) = 1 - \rho_2^2 \neq 0$. If either (i) or (ii) is violated, one 
recovers level repulsion $P(S\rightarrow 0)\rightarrow 0$.
Indeed, numerous numerical studies (\cite{PR94b,PR97,P95} and references
therein) give phenomenological support to the {\em fractional power-law 
level repulsion} which is usually very well globally captured by the
phenomenological Brody distribution \cite{B73} 
\begin{equation}
P_{\rm B}(S) = (\beta+1)b S^\beta \exp(-b S^{\beta+1}),\quad 
b = [\Gamma(1+(\beta+1)^{-1})]^{\beta+1}
\label{eq:Brody}
\end{equation}
In fact, even for a generic 2-dim toy system with a simple phase space
structure (where (i) and (ii) have the largest chances to apply)
having a small number of islands and well connected chaotic component, 
one may verify that (i) and (ii) are typically fulfilled only for 
sequential quantum numbers substantially larger than $\sim 10^6-10^7$ 
\cite{P96}.

So it is not surprising that the `ultimate semiclassical' BR statistics 
have so far been clearly demonstrated only in two toy systems: 
(1) in a rather abstract compactified standard map \cite{PR94a}, and (2) 
in a 2-dim semi-separable oscillator \cite{P95,P96}, which is dynamically
a generic system but geometrically somewhat special. 
Here we give the first clear numerical demonstration of BR statistics 
in a generic billiard system with a smooth boundary.
We consider classical and quantum motion of a free particle moving
inside a bounded planar region which has a shape of a smoothly deformed
circle. Billiard domain is described by the following function $r(\phi)$, 
giving the radial distance from the origin to the boundary as a function 
of the polar angle $\phi$,
\begin{equation}
r(\phi) = 1 + a\cos(4\phi).
\label{eq:bil}
\end{equation}
For the purpose of this letter we choose the following value of
deformation parameter, $a=0.04$, for which the
classical phase space (plotted in a Poincar\' e-Birkhoff coordinates on
a boundary-section in figure 1) has regular regions with the 
total relative Liouville measure (not the area on SOS \cite{Meyer})
$\rho_1^{\rm cl} = 0.115 \pm 0.005$. Note that numerical computation of 
measures of regular and chaotic components of phase space in mixed (KAM) systems 
converges very slowly with increasing discretization of the phase space 
\cite{Dobnikar}, hence it is difficult to further reduce the error estimate 
$\delta \rho_1^{\rm cl}\approx 0.005$.

High-lying quantum eigenenergies, eigenvalues of the Schr\" odinger
equation $(\nabla^2 + k^2)\Psi_k(\vec{r}) = 0$ with Dirichlet b.c. 
on the boundary $r=r(\phi)$, have been computed by means of extremely efficient
scaling technique proposed by Vergini and Saraceno \cite{VS95}:
Eigenstates $\Psi_k$ are expanded in a basis of {\em circular scaling
functions} (see also \cite{CP}, 
as opposed to plane waves used in original approach \cite{VS95})
\begin{equation}
\Psi_k(\vec{r}) = \sum_{l=1}^{M} a_l J_{4l}(kr) \sin(4 l \phi).
\end{equation}
Note that the billiard has been desymmetrized and here we
consider only fully antisymmetric states with
respect to 8-fold symmetry group of the billiard.
The coefficients $a_l$ are determined by minimizing a
special positive quadratic form defined along the boundary
of the billiard \cite{VS95}. 
The dimension of the problem $M=[(1+a)k/4] + M_{\rm evanescent}$
is nearly optimal where few ten, typically $M_{\rm evanescent}\sim 40$, 
evanescent modes have been added in order to ensure convergence and 
accuracy of the computed energy levels.
We should note that the scaling method of quantization of
billiards is by far superior to other relevant methods, e.g. 
boundary integral method \cite{BerryWilk84} or Heller's 
plane wave decomposition \cite{Heller}, since it yields a constant
fraction ($5\%-10\%$) of $M\propto k$ of accurate levels, with no
risk of missing any,
by solving a single generalized eigenvalue problem of dimension $M$.

In figure 2 we show cumulative nearest neighbour level spacing
distribution $W(S)=\int_0^S ds P(s)=(d/dS)E(S)-(d/dS)E(0)$ for the 
{\em unfolded} \cite{H91} spectral stretches 
$\{ e_n = k_n^2/32 + (1/8+1/\pi)k_n; 
k_{\rm min} \le k_n \le k_{\rm max}\}$ (for small $a$) 
containing about $6000$ consecutive levels each.
In fact, we have computed several spectral stretches, the first in 
the {\em near semiclassical regime} $399.7 \le k \le 600.1$ 
(containing 6220 levels), and the last in the {\em far semiclassical regime} 
$15999.707 \le k \le 16004.865$ (containing 5168 levels) where the {\em sequential 
quantum number} is $N\approx k^2/32 + (1/8+1/\pi)k \approx 0.8\cdot 10^7$.
Only for the last spectral stretch in the far semiclassical regime 
($k\approx 16 000$) we found statistically significant agreement with BR distribution (figures 2,3) where the quantal 
(best-fitting) parameter $\rho_1^{\rm q}$ agrees very well with its 
classical value, namely $\rho_1^{\rm q} = 0.119$.
However, for smaller sequential quantum numbers, when we approach 
the near-semiclassical regime, we find substantial deviation from BR statistics
and recover {\em fractional-power law level repulsion} \cite{PR93,PR94b}, 
namely for the lowest spectral stretch (figures 2,3) at $k\approx 500$ we find 
almost statistically significant agreement with Brody distribution 
(\ref{eq:Brody}) with exponent $\beta = 0.46$.
Of course, the fit to BR distribution in the near semiclassical regime 
$k\approx 500$ and the fit to Brody distribution in the far semiclassical 
regime $k\approx 16000$ turned out to be highly statistically 
{\em non-significant}.

In figure 3 we show deviations of numerical spacing
distributions from the semiclassical BR distribution
(for parameter $\rho^{\rm q}_1=0.119\approx\rho^{\rm cl}_1$)
in fine detail, using a smooth U-transformation \cite{PR93}
of the cumulative level spacing distribution $U(W(S))-U(W_{\rm BR}(S))$, 
where $U(W) = (2/\pi)\arccos\sqrt{1-W}$, against $W(S)$. This
statistical representation has a uniform expected statistical error 
$\delta U(W) = 1/(\pi\sqrt{\Delta N})$ (where $\Delta N$ is the number 
of levels in a spectral stretch) and a constant density of numerical 
points along the abscissa.
One can see very clearly that in both cases, far and near semiclassical,
the numerical distributions are fluctuating around theoretical
BR and Brody distributions, respectively, within expected statistical error.

Finally we wish to characterize long-range spectral correlations 
as well, so we consider the number variance 
$\Sigma^2(L) = \ave{N^2}_L-\ave{N}^2_L$, i.e. the variance
of the number of unfolded levels $e_n$ in an interval of length 
$L$. Since this is a linear statistic it should be additive upon
statistically independent superposition of spectral subsequences \cite{SV}.
According to assumptions (i) and (ii) one immediately arrives to 
the ultimate semiclassical formula for the number variance \cite{SV}
\begin{equation}
\Sigma^2(L) = \Sigma^2_{\rm Poisson}(\rho_1 L) + \Sigma^2_{\rm GOE}(\rho_2 L)
\label{eq:SVF}
\end{equation}
where $\Sigma^2_{\rm Poisson}(L) = L$ is the number variance of Poissonian
level sequence, and $\Sigma^2_{\rm GOE}(L)\approx (2/\pi^2)\ln(2\pi L)$ 
is the number variance of the spectrum of infinitely dimensional GOE random
matrix which is supposed to model chaotic levels.
In figure 4 we show $\Sigma^2(L)$ for four spectral stretches,
namely for $k\approx 500$, $k\approx 2000$, $k\approx 8000$, 
and $k\approx 16000$, and only the last in the far semiclassical regime 
agrees well with the formula (\ref{eq:SVF}) (for parameter 
$\rho_1=\rho_1^{\rm cl} = 0.115$) up to $L=L^*\approx 50$.
\\\\
In this letter we have clearly demonstrated the validity of BR level spacing 
distribution in a generic smooth plane billiard system with mixed classical 
phase space, namely the quartic billiard.
However, for insufficiently small semi-classical parameter $\hbar\sim N^{-1/2}$,
we demonstrated the existence of fractional-power law level
repulsion which is (for sufficiently small energy ranges) globally very well 
captured by the phenomenological Brody distribution. Unfortunately, this is the 
regime which can only be observed in most experimental situations due to 
extremely high energy region of crossover to BR statistics.

We should note that this particular KAM billiard system ((\ref{eq:bil}) for 
$a=0.04$) has quite simple phase space structure which is reflected in 
relatively low transition point ($N\approx 10^7$) to the ultimate semiclassical 
BR statistics. For example, in a well known quadratic or Robnik billiard, the 
phase space is much more complicated \cite{R83} 
(smaller regular islands, more partial phase space bariers, cantori), 
and as a consequence, the transition to BR 
regime is shifted to much higher energies \cite{RLV}.

\section*{Acknowledgments}

Discussions and collaboration on related projects with Marko Robnik, as well 
as the financial support from the Ministry of Science and Technology of R 
Slovenia are gratefully acknowledged.

\vfill
\newpage

\section*{Figure captions}
\bigskip
\bigskip

\begin{figure}[htbp]
\hbox{
\leavevmode
\epsfxsize=5in
\epsfbox{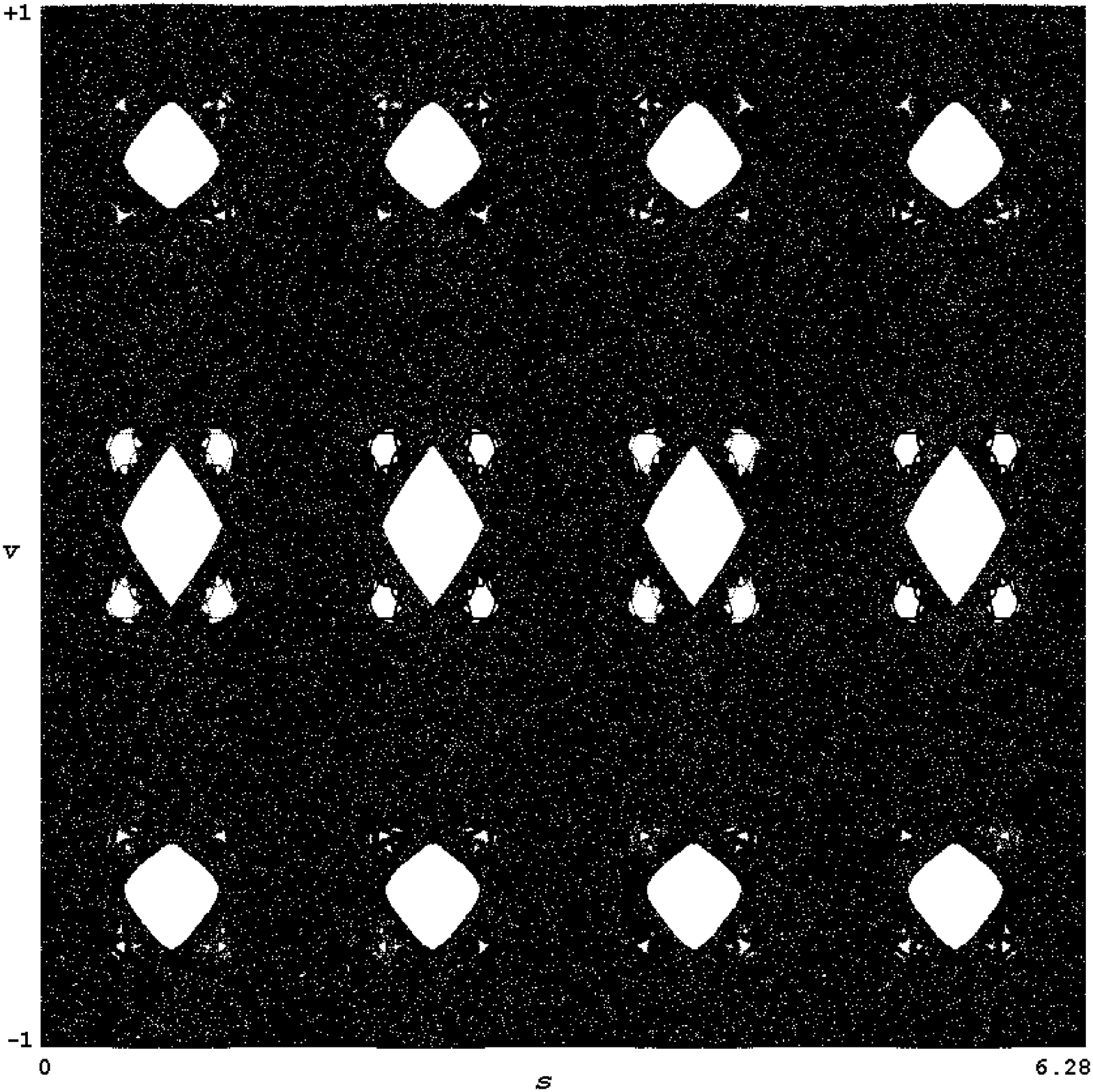}}
\caption{
Classical phase space for $a=0.04$ in Poincar\' e-Birkhoff coordinates: arc-length
$s$ and tangential (normalized) velocity $v$.
We show a chaotic orbit with $2 000 000$ collisions with the boundary.
}
\label{fig:1}
\end{figure}

\begin{figure}[htbp]
\hbox{\hspace{-1.5in}\vbox{
\hbox{
\leavevmode
\epsfxsize=8in
\epsfbox{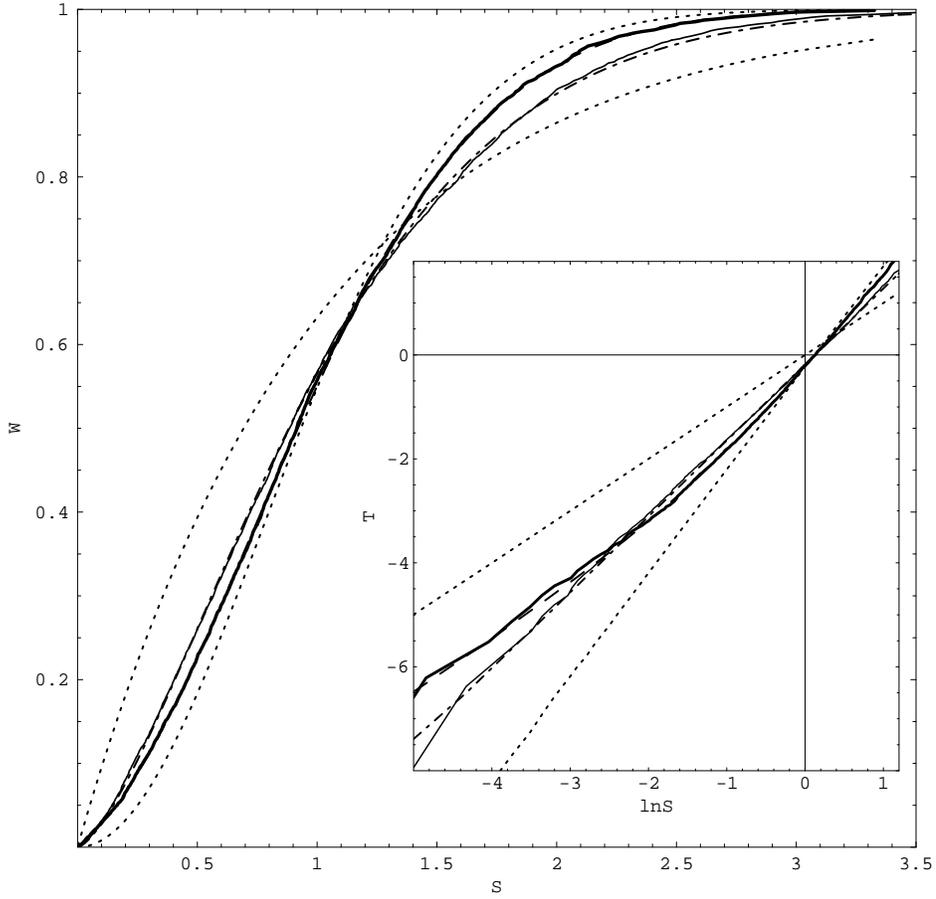}}
}}
\caption{
Cumulative nearest level spacing distribution $W(S)$ for a
stretch of 5168 consecutive levels in the
far semiclassical regime ($k\approx 16 000$) (thick curve)
and a stretch of 6220 consecutive levels in the near semiclassical
regime ($k\approx 500$) (thin curve).
The first numerical curve is almost overlapping with theoretical best fitting
BR  distribution for $\rho_1^{\rm q}=0.119$ (dashed curve),
while the second numerical curve agrees very well with the best fitting Brody
distribution with exponent $\beta=0.46$ (dot-dashed curve).
For comparison we give Poisson and GOE integrated level spacing
distributions (dotted curves).
In the inset we plot the same data in the T-function 
representation \cite{PR93}, $T(S) = \ln(-\ln(1-W(S)))$
against $\ln S$, which transforms the Brody distributions (an hence also
Poissonian and Wigner) to straight lines, and enhances the region of
small spacings.
}
\label{fig:2}
\end{figure}

\begin{figure}[htbp]
\hbox{
\leavevmode
\epsfxsize=5in
\epsfbox{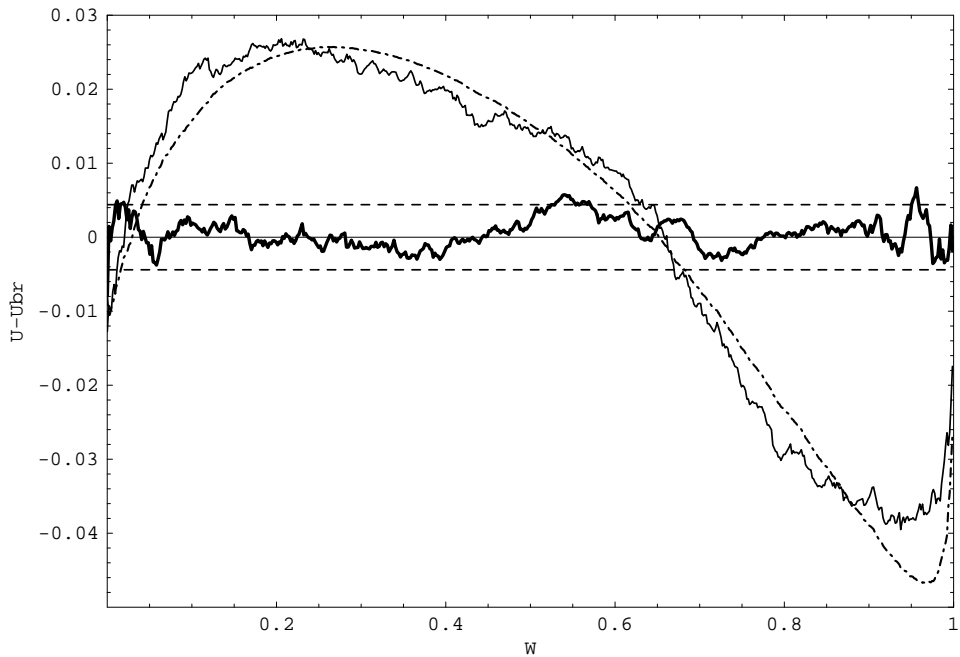}}
\caption{
Fine detail deviations from Berry-Robnik distribution (for $\rho_1=0.119$) 
in a uniform U-function transformation \cite{PR93}: we plot
$U(W(S))-U(W_{\rm BR}(S))$ against $W(S)$.
In the far semiclassical regime $k\approx 16 000$ (5168 consecutive levels),
the difference of U-functions (thick curve) lies within a band of expected 
statistical error $\delta U$ (dashed lines), while in
the near semiclassical regime $k\approx 500$ (6220 consecutive levels),
the difference of U-functions (thin curve) agrees very well with
the difference of U-functions for the best fitting Brody distribution 
with exponent $\beta=0.46$ (dash-dotted curve).
}
\label{fig:3}
\end{figure}

\begin{figure}[htbp]
\hbox{
\leavevmode
\epsfxsize=5in
\epsfbox{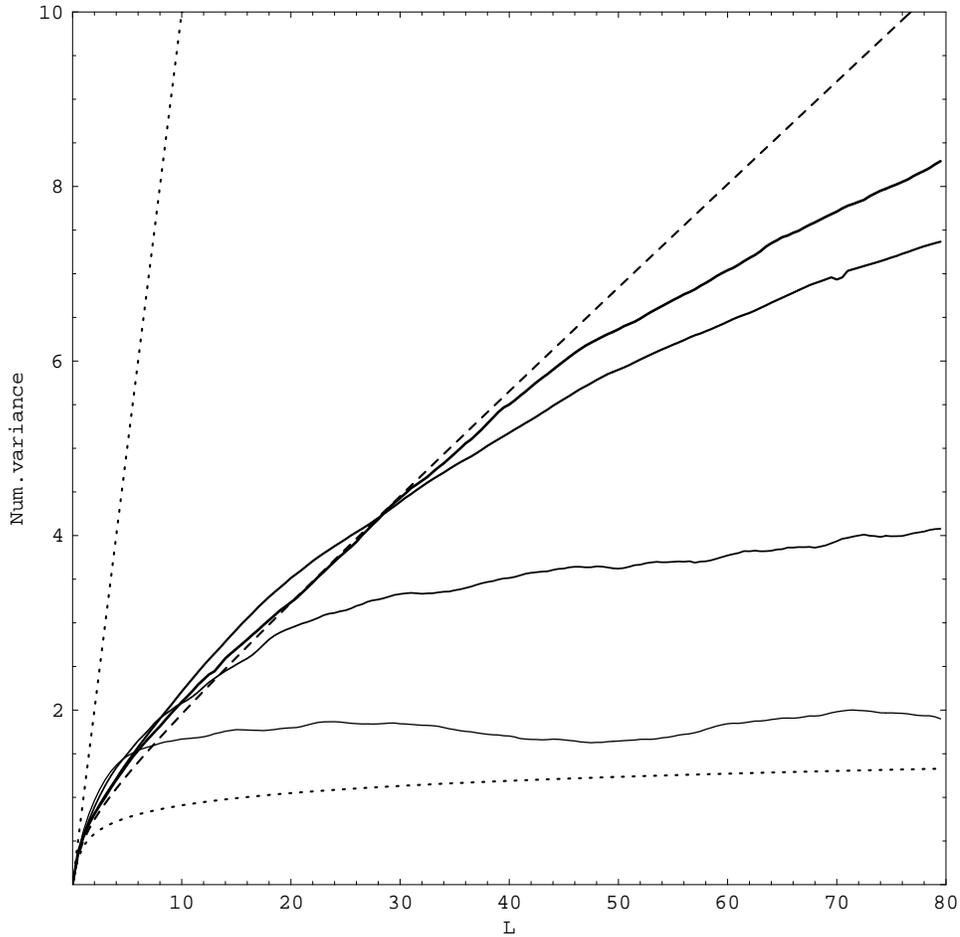}}
\caption{
Number variance $\Sigma^2(L)$ for the four spectral stretches: for 
$k\approx 16 000$ (thickest curve, 5168 levels), 
$k\approx 8000$ (next thickest curve, 17300 levels),
$k\approx 2000$ (second thinnest curve, 5100 levels), 
and $k\approx 500$ (thinnest curve, 6220 levels).
Dashed curve is the semiclassical formula (\ref{eq:SVF}) which
indeed reproduces the far semiclassical numerical data (thickest full 
curve) quite well, for $L\le L^* \approx 50$.
For comparison we give the Poissonian and GOE curves (dotted).
}
\label{fig:4}
\end{figure}

\end{document}